\documentclass[aps,twocolumn,floatfix]{revtex4-1}
\usepackage{bm} 
\usepackage{dcolumn} 
\usepackage{graphicx} 
\usepackage{amsmath} 
\usepackage{amssymb} 
\usepackage{amsfonts} 

\begin{document}
\title{Electrodynamics on Fermi Cyclides in Nodal Line Semimetals}

\author{Seongjin Ahn$^{1}$}
\author{E. J. Mele$^2$}
\email{mele@physics.upenn.edu}
\author{Hongki Min$^{1,2}$}
\email{hmin@snu.ac.kr}
\affiliation{$^1$ Department of Physics and Astronomy, Seoul National University, Seoul 08826, Korea}
\affiliation{$^2$
Department of Physics and Astronomy, University of Pennsylvania, Philadelphia, Pennsylvania 19104, USA}

\date{\today}

\begin{abstract}
We study the frequency-dependent conductivity of nodal-line semimetals (NLSMs)
 focusing on the effects of carrier density and energy dispersion on the nodal line. We find that the low-frequency conductivity has a rich spectral structure which can be understood using scaling rules derived from the geometry of their Dupin cyclide Fermi surfaces. We identify different frequency regimes, find scaling rules for the optical conductivity in each and demonstrate them with numerical calculations of the inter- and intraband contributions to the optical conductivity using a low-energy model for a generic NLSM.
 \end{abstract}

\maketitle

{\em Introduction.} ---
Interest in topological states of quantum matter has led to the identification of new gapless electronic states that support nontrivial geometric structures in their band structures at or near the Fermi energy. In topological semimetals, the conduction and valence bands contact at points or lines in  momentum space, and the band degeneracy at the contact is protected by symmetries such as crystalline and time-reversal symmetries.  Among the three-dimensional topological semimetals, Weyl (Dirac) semimetals have nondegenerate (degenerate) conduction and valence bands which touch at discrete points, whereas in nodal-line semimetals (NLSMs), the two bands cross on closed lines in momentum space \cite{Burkov2011,Chiu2014,Fang2016,Rui2017}.

\begin{figure}[!htb]
\includegraphics[width=0.9\linewidth]{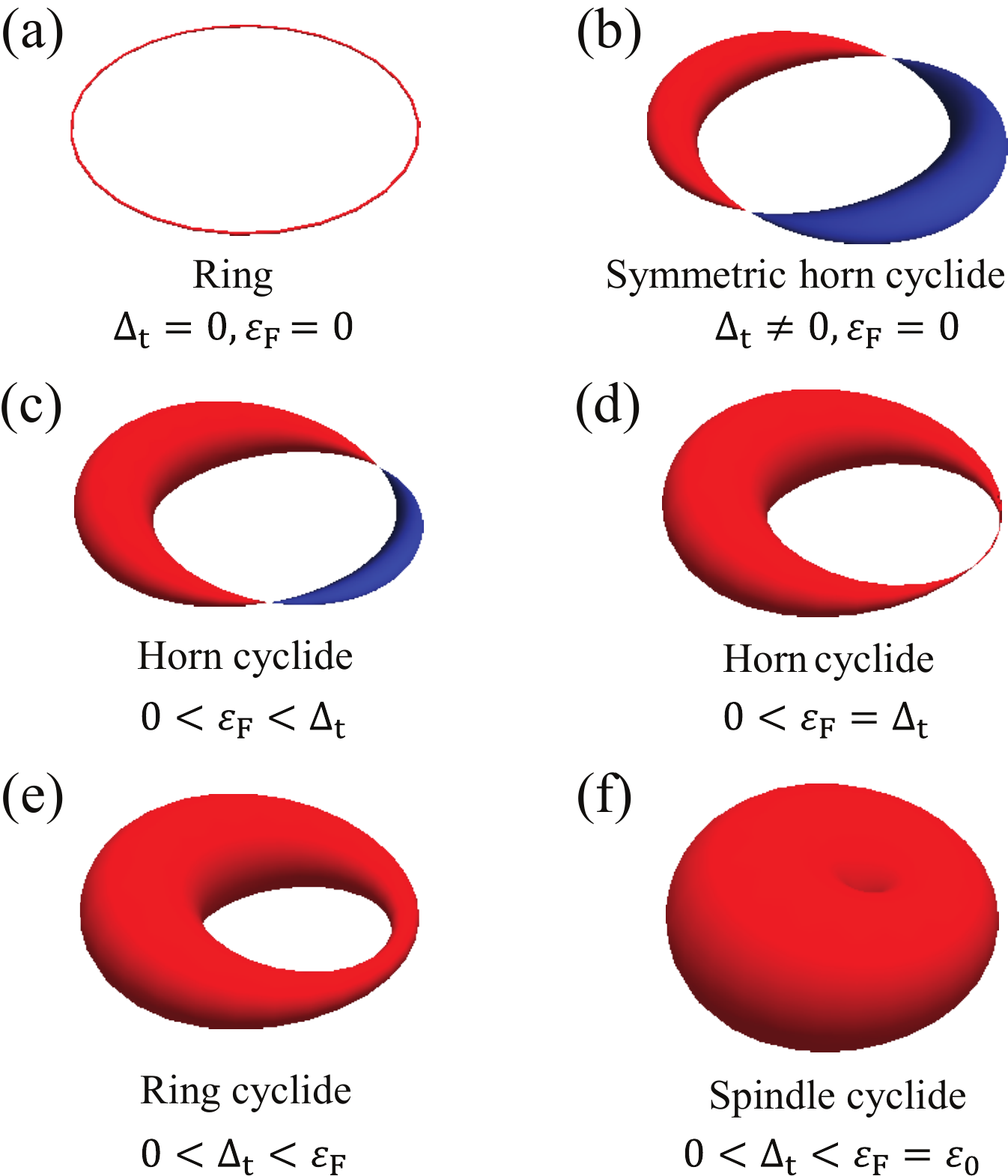}
\caption{
Evolution of the Fermi surface (FS) as a function of tilt $\Delta_{\rm t}$ and Fermi energy $\varepsilon_{\rm F}$ with red (blue) indicating the electron (hole) pocket.
(a) At zero Fermi energy with zero tilt, the FS has a one-dimensional ring shape sitting on the zero-energy plane. (b) With a finite tilt, the ring shape evolves into a symmetric horn cyclide containing both electron and hole pockets symmetrically, which vanish at two contact points. Upon increasing the Fermi energy, the electron and hole pockets (c) become asymmetric and the contact points move out from the symmetrical axis, (d) converge into a single point when the Fermi energy equals the tilt energy, and (e) vanish when the Fermi energy becomes larger than the tilt energy. (f) When the Fermi energy equals the energy scale of the ring radius, the FS is merged into a spherelike shape with no holes in the center, similar to that of Weyl semimetals.
}
\label{fig:duplin_cyclide}
\end{figure}

When protected by a mirror symmetry, such a nodal line is ``flat" in the sense that it lies in a single plane in $\bm{k}$ space. However, the contact line is not similarly constrained to occur on a constant energy surface. An energy dispersion on the nodal line has no effect on its topological character, which is determined by the phase winding of the Bloch states around the singularity. However it generically forces the system into a semimetallic state with coexisting electron and hole pockets and an unconventional Fermi surface (FS) geometry, exhibiting the rich structures of Dupin cyclide geometries \cite{Eisenhart}, as shown in Fig.~\ref{fig:duplin_cyclide}. 
Here the geometry of the FS is determined by a combination of the energy dispersion of the contact line (tilt) and the Fermi energy which play a crucial role in determining various physical properties in NLSMs.

The optical conductivity of a low-energy model for a NLSM in the absence of tilt has recently been studied \cite{Carbotte2017}. However, once an energy tilt is introduced, there is a competition between two energy scales set by the amount of dispersion and by the chemical potential.
In this work, we study the consequences of this competition for the low-frequency conductivity of a NLSM and analyze its characteristic frequency dependence using the geometry of the Dupin cyclide. We find new spectral features that occur as a result of its unconventional geometry. For a nonzero Fermi energy smaller than the tilt energy scale, full Pauli blocking is prevented and instead all three diagonal components of the optical conductivity tensor show linear scaling with frequency. For the Fermi energy larger than the tilt energy, the interband optical conductivity recovers a gap due to Pauli blocking. We find nonanalytic features in both the frequency dependence of the interband conductivity and the chemical potential dependence of the Drude stiffness which arises from Lifshitz transitions of the FS.


{\em Model.} --- In the continuum approximation, the minimal Hamiltonian for tilted NLSMs that captures the essential features of its low-energy excitations takes the form of a 2 by 2 matrix given by
\cite{Mullen2015,Fang2015}
\begin{equation}\label{eq:hamiltonian_2band}
H=\hbar v q_\rho \sigma_x +\hbar v k_z \sigma_y + \hbar \bm{v}_{\rm t}\cdot\bm{k}\sigma_0,
\end{equation}
where $\sigma_x$ and $\sigma_y$ are the Pauli matrices, $\sigma_0$ is the identity matrix, $\bm{v}_{\rm t}$ is the tilt velocity, $q_{\rho}=k_{\rho}-k_0$, $k_{\rho}=\sqrt{k_x^2+k_y^2}$, and $k_0$ is the radius of the nodal ring.
The eigenenergies of the Hamiltonian are given by
\begin{equation}
\varepsilon_{\pm}(\bm{k})=\pm \hbar v\sqrt{q_{\rho}^2 + k_z^2} + \hbar \bm{v}_{\rm t}\cdot \bm{k},
\end{equation}
which has a ring shape zero-energy contour with a slope of $\bm{v}_{\rm t}$.
In the following, we consider tilt only in the in-plane direction because it produces electron-hole pockets, leading to qualitative changes in the optical conductivity, whereas tilt along the perpendicular axis has little effect on the optical conductivity unless it is so extreme that the system is in a type-II semimetallic state \cite{supplemental}. Without the loss of generality, we therefore set
$\bm{v}_{\rm t}=v_{\rm t} \hat{x}$.

\begin{figure}[!htb]
\includegraphics[width=1\linewidth]{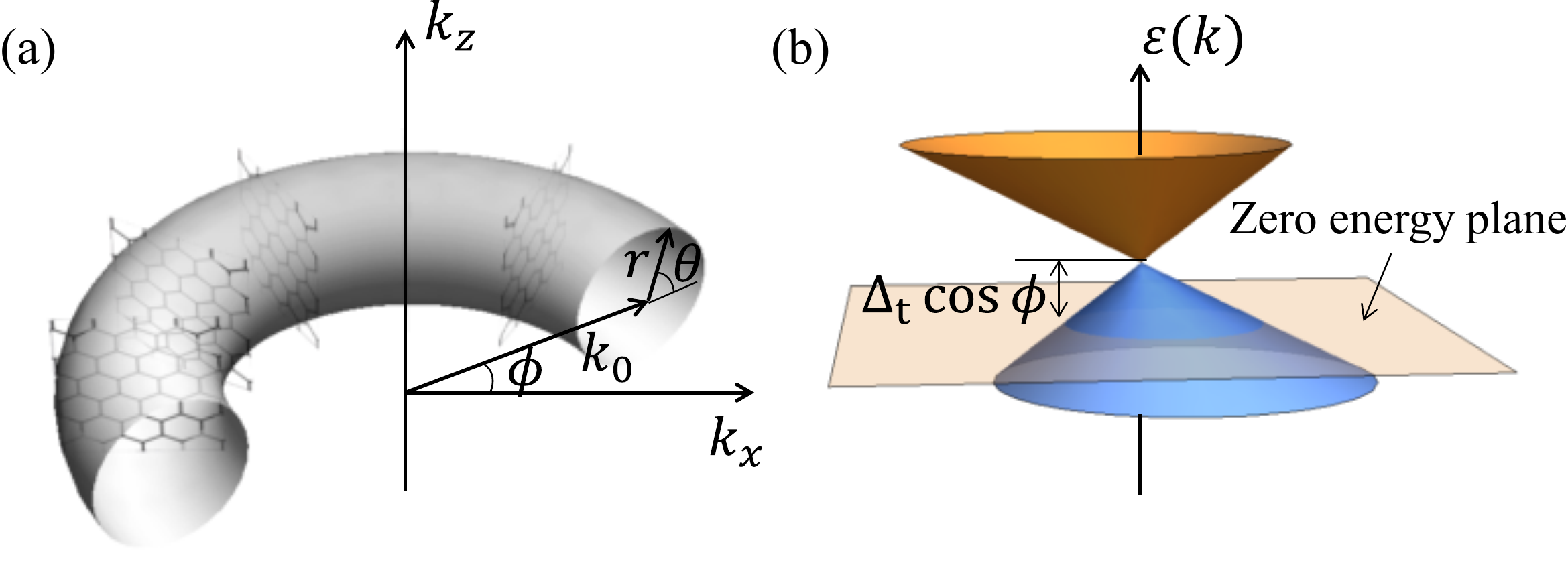}
\caption{
(a) Schematic illustration of toroidal coordinates and graphene sheets standing perpendicular to the nodal-line plane. (b) The Dirac cone energy dispersion of the graphene sheet located at $\phi$. Note that the Dirac cone is shifted from the zero energy by $\Delta_{\rm t}\cos{\phi}$.}
\label{fig:torus}
\end{figure}

 Remarkably, in toroidal coordinates \cite{supplemental}, this NLSM Hamiltonian in Eq.~(\ref{eq:hamiltonian_2band}) can be written in the same form as the low-energy graphene Hamiltonian, but expressed in polar coordinates ($r$, $\theta$) centered at $\bm{k}_0$ [see Fig.~\ref{fig:torus}(a)]:
\begin{equation}
\label{eq:hamiltonian_graphene}
H =
\varepsilon_0 \left(
\begin{array}{cc}
0 & \tilde{r}e^{-i \theta} \\
\tilde{r}e^{i \theta} & 0
\end{array}
\right)
+\Delta_{\rm t}(1+\tilde{r} \cos{\theta})\cos{\phi} \sigma_0,
\end{equation}
where $\tilde{r}=r/k_0$, $\varepsilon_0=\hbar v k_0$, and $\Delta_{\rm t}=\hbar v_{\rm t} k_0$.
Thus, in the energy range where the toroidal structure is maintained, we can consider NLSMs as a collection of graphene sheets with tilt in the $x$ direction and the Dirac point shifted from zero energy by $\Delta_{\rm t} \cos{\phi}$ [see Fig.~\ref{fig:torus}(b)].

{\em Optical conductivity obtained from a collection of graphene sheets.} ---
In the linear response limit,
we obtain the optical conductivity of NLSMs by summing up individual contributions from each of the graphene sheets:
\begin{equation}
\label{eq:NLSMs_optical_conductivity}
\sigma_{ii}^{\rm NLSM}
=k_0\int^{2\pi}_{0}\frac{d\phi}{2\pi}\; \sigma^{\rm gr}(\phi) \mathcal{F}_{ii}(\phi),
\end{equation}
where $i=x,y,z$, $\sigma^{\rm gr}(\phi)$ is the optical conductivity of a graphene sheet located at $\phi$ with tilt in the $x$ direction,
and $\mathcal{F}_{xx}(\phi)=\cos^2{\phi}$,
$\mathcal{F}_{yy}(\phi)=\sin^2{\phi}$, $\mathcal{F}_{zz}(\phi)=1$ are geometric factors from the projection of an external electric field on the graphene sheet and that of the in-plane velocity of graphene on the current direction.
Note that Eq.~(\ref{eq:NLSMs_optical_conductivity}) is valid for $\hbar\omega<2\varepsilon_0$ where the toroidal structure is maintained.

First, consider the case of $\Delta_{\rm t}=0$. Since the optical conductivity from a single Dirac cone filled to energy $\varepsilon_{\rm F}$ is given by $\sigma^{\rm gr}(\phi)=\frac{e^2}{16\hbar}\Theta\left(\hbar\omega-2|\varepsilon_{\rm F}|\right)$ \cite{Ando2002}, the optical conductivity of the NLSM is
\begin{equation}
\label{eq:NLSMs_graphene_optical_conductivity}
{\sigma_{ii} \over \sigma_0}\approx
\begin{cases}
\frac{1}{2}\Theta\left(\hbar\omega-2|\varepsilon_{\rm F}|\right) &\text{for $i=x,y$},\\
\Theta\left(\hbar\omega-2|\varepsilon_{\rm F}|\right) &\text{for $i=z$},
\end{cases}
\end{equation}
where $\sigma_0={e^2 k_0 \over 16\hbar}$.
Thus, for $\hbar\omega<2|\varepsilon_{\rm F}|$ the optical conductivity vanishes due to Pauli blocking, whereas for $2|\varepsilon_{\rm F}|<\hbar\omega<2\varepsilon_0$, it remains constant \cite{Carbotte2017}.

For $\Delta_{\rm t}\neq0$, the conductivity should be modified to take into account (i) the shift of the Dirac point from zero energy [first term of $\sigma_0$ in Eq.~(\ref{eq:hamiltonian_graphene})] and (ii) the tilted linear band dispersion (second term). At low frequencies ($\hbar\omega\ll \varepsilon_0$), however, when the Dirac points lie close to zero energy, the tilt in the band dispersion is negligible. Thus, $\sigma_{ii}(\omega)$ at low frequencies can be obtained using $\sigma^{\rm gr}(\phi)\approx\frac{e^2}{16\hbar}\Theta\left(\hbar\omega-2|\varepsilon_{\rm F}-\Delta_{\rm t}\cos\phi)|\right)$ in Eq.~(\ref{eq:NLSMs_optical_conductivity}), giving
\begin{equation}
\label{eq:NLSMs_optical_conductivity_normalized}
{\sigma_{ii} \over \sigma_0}
\approx\int_{0}^{2\pi} {d\phi\over 2\pi}\,
\Theta({\hbar\omega-2\left|\varepsilon_{\rm F}-\Delta_{\rm t}\cos{\phi}\right|}) {\mathcal F}_{ii}(\phi).
\end{equation}

In the following, we present numerically calculated optical conductivities over a wide frequency range in the presence of tilt and finite Fermi energy by evaluating the Kubo formula [see Eq.~(6) in Supplemental Material \cite{supplemental}]. We analyze the results by investigating the geometry of the phase space (PS) for interband transitions and the low-frequency analytic forms obtained from Eq.~(\ref{eq:NLSMs_optical_conductivity_normalized}).
Here, the PS for interband transitions is given by the intersection between the outside of the FS and the PS allowed by energy conservation.
For a given frequency $\omega$, the PS for NLSMs allowed by energy conservation is the surface of the momentum space torus which satisfies $\hbar\omega=\varepsilon_{+}(\bm{k})-\varepsilon_{-}(\bm{k})$.

\begin{figure}[!htb]
\includegraphics[width=1\linewidth]{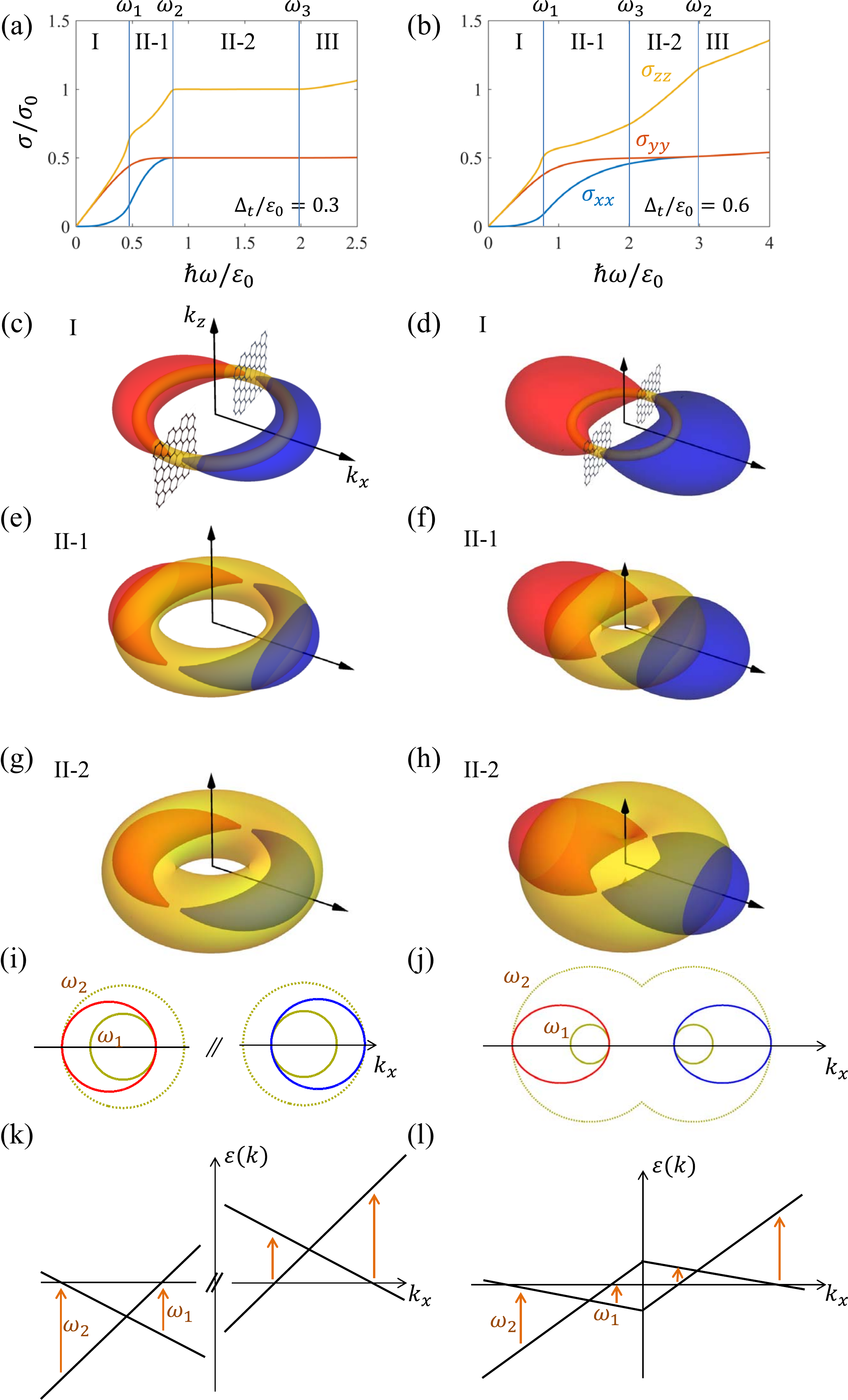}
\caption{
(a), (b) Calculated optical conductivities of NLSMs for $\varepsilon_{\rm F}=0$ with two different tilt energies of (a) $\Delta_{\rm t}=0.3\varepsilon_0$ and (b) $\Delta_{\rm t}=0.6\varepsilon_0$. Regions I, II-1, II-2, III represent the frequency domains in which the PS for interband transitions grows continuously without any abrupt changes.
(c)-(h) The PS allowed by energy conservation indicated by a yellow torus along with electron (red) and hole (blue) pockets in different frequency domains. Note that in region I the PS allowed for interband transitions consists of two local domains, while in region II they merge together to form a connected geometry.
(i), (j) Cross-sectional views of the electron-hole pockets and the PS allowed by energy conservation in the $k_x-k_z$ plane at the frequencies of $\omega=\omega_1$ (yellow solid lines) and $\omega=\omega_2$ (yellow dashed lines), where the PS allowed for interband transitions changes its geometry leading to kink structures in the optical conductivity. (k), (l) Energy band dispersions along the $k_x$ axis with $k_y=k_z=0$ for (k) $\Delta_{\rm t}=0.3\varepsilon_0$ and (l) $\Delta_{\rm t}=0.6\varepsilon_0$. The geometrical changes occur at frequencies corresponding to the onset of interband transitions indicated by arrows.
}
\label{fig:optical_conductivity_tilt}
\end{figure}

{\em Optical conductivity with $\varepsilon_{\rm F}=0$} ---
We first consider the case where $\varepsilon_{\rm F}=0$ in the presence of tilt.
Figures \ref{fig:optical_conductivity_tilt}(a) and \ref{fig:optical_conductivity_tilt}(b) show calculated optical conductivities with tilt energies of $\Delta_{\rm t}=0.3\varepsilon_0$ and $\Delta_{\rm t}=0.6\varepsilon_0$, respectively. Note that the optical conductivities show kink structures at transitions between different frequency domains (I, II and III) and characteristic frequency dependences determined by the tilt energies.

In region I, the optical conductivity in the tilt direction is $\sigma_{xx} \propto \omega^3$ at low frequencies while those in the other directions ($\sigma_{yy}$ and $\sigma_{zz}$) are $\propto \omega$ [see region I in Figs.~\ref{fig:optical_conductivity_tilt}(a) and \ref{fig:optical_conductivity_tilt}(b)].
Figures \ref{fig:optical_conductivity_tilt}(c) and \ref{fig:optical_conductivity_tilt}(d) show the corresponding PS for interband transitions, which is divided into two separated islands located at $\phi=\pm\frac{\pi}{2}$.
As demonstrated above in Eq.~(\ref{eq:NLSMs_optical_conductivity_normalized}), we can express the optical conductivity by averaging contributions from the graphene sheets that occupy these two isolated regions. In the low-frequency limit, in the vicinity of the contact points, the optical conductivity is approximately given by
\begin{eqnarray}
\label{eq:NLSMs_optical_conductivity_notilt}
{\sigma_{ii} \over \sigma_0}
&\approx&
\int_{\phi\sim\pm\frac{\pi}{2}} \frac{d\phi}{2\pi}\; \Theta({\hbar\omega-2\left|\Delta_{\rm t}\cos{\phi}\right|}) \mathcal{F}_{ii}(\phi)\nonumber\\
&\approx& \frac{2}{\pi} \int^{\frac{\hbar\omega}{2\Delta_{\rm t}}}_{0} d\phi\; \mathcal{F}_{ii}\left(\phi+\frac{\pi}{2}\right).
\end{eqnarray}
With the expansion of $\mathcal{F}_{ii}\left(\phi+\frac{\pi}{2}\right)$
around $\phi=0$, the resulting conductivities in the lowest order are
\begin{equation}
\label{eq:NLSMs_optical_conductivity_noEf}
{\sigma_{ii} \over \sigma_0}=
\begin{cases}
\frac{1}{12\pi}\left(\frac{\hbar\omega}{\Delta_{\rm t}}\right)^3 &\text{for $i=x$},\\
\frac{\hbar\omega}{\pi\Delta_{\rm t}} &\text{for $i=y,z$},
\end{cases}
\end{equation}
which agree with our numerical results.

At the intersection between regions I and II-1, the PS allowed by energy conservation (yellow torus) begins to touch the boundary of electron (red) and hole (blue) pockets, and the two isolated PS regions for interband transitions merge forming a connected geometry distinguished from that in region I [see Figs.~\ref{fig:optical_conductivity_tilt}(e) and \ref{fig:optical_conductivity_tilt}(f)].
Note that this change of geometry produces a kink in $\sigma_{ii}$ seen most clearly in $\sigma_{zz}$ because of its $\phi$-independent projection factor [$\mathcal{F}_{zz}(\phi)=1$; see Eq.~(\ref{eq:NLSMs_optical_conductivity})].
Such a geometrical change also occurs at the intersection between II-1 and II-2 regions giving a kink in $\sigma_{zz}$. By observing the cross-sectional view of the PS allowed by energy conservation and the FS in the $k_x-k_z$ plane [see Figs.~\ref{fig:optical_conductivity_tilt}(i) and ~\ref{fig:optical_conductivity_tilt}(j)], we can calculate the frequencies at which kinks appear. Alternatively, since the geometrical changes associated with additional interband transitions occur along the $k_x$ axis, the problem of finding kink frequencies is reduced to obtaining $\omega_1$ and $\omega_2$ in Figs.~\ref{fig:optical_conductivity_tilt}(k) and ~\ref{fig:optical_conductivity_tilt}(l): $\hbar\omega_1=\frac{2\Delta_{\rm t}}{\varepsilon_0 + \Delta_{\rm t}}\varepsilon_0$, $\hbar\omega_2=\frac{2\Delta_{\rm t}}{\varepsilon_0 - \Delta_{\rm t}}\varepsilon_0$.

In region II-2, for $\Delta_{\rm t}=0.3\varepsilon_0$, the PS allowed by energy conservation [yellow torus in Fig.~\ref{fig:optical_conductivity_tilt}(g)] covers the whole FS while keeping its toroidal structure similar to the untilted case, thus the optical conductivity shows flat behavior with {\it exactly the same height as that of untilted NLSMs}. For $\Delta_{\rm t}=0.6\varepsilon_0$, however, the PS allowed by energy conservation [yellow spherelike manifold in Fig.~\ref{fig:optical_conductivity_tilt}.(h)] is no longer a torus and does not fully cover the whole FS, exhibiting a monotonic increase in $\sigma_{zz}$ instead of the flat behavior. Note that the PS allowed by energy conservation changes its geometry from a torus to a spherelike manifold at the frequency $\hbar\omega_3=2\varepsilon_0$. Thus, the condition for the existence (absence) of the flat region can be obtained from the condition $\omega_2<\omega_3$ ($\omega_2>\omega_3$), leading to $\Delta_{\rm t}<0.5 \varepsilon_0$ ($\Delta_{\rm t}>0.5 \varepsilon_0$).


In region III, the PS allowed by energy conservation for both $\Delta_{\rm t}=0.3\varepsilon_0$ and $\Delta_{\rm t}=0.6\varepsilon_0$ merge into a sphere-like geometry covering the whole FS, similar to that of Weyl semimetals. Thus, at high frequencies in this frequency domain, the optical conductivity shows a linear behavior, as already shown in previous studies \cite{Carbotte2017}.

\begin{figure}[!htb]
\includegraphics[width=1\linewidth]{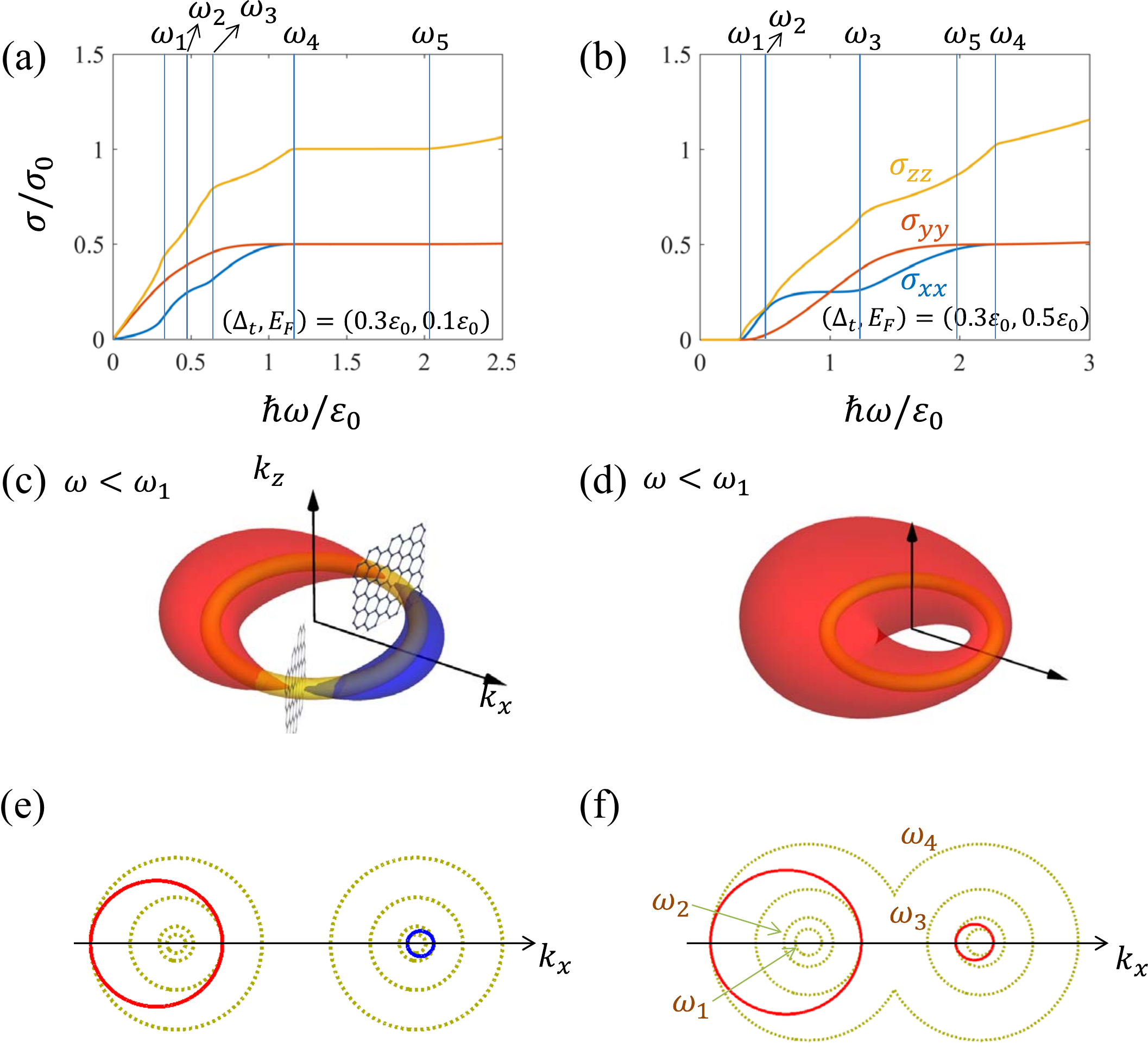}
\caption{
(a), (b) Calculated optical conductivities of tilted NLSMs with $\Delta_{\rm t}=0.3\varepsilon_0$ for two different Fermi energies of (a) $\varepsilon_{\rm F}=0.1\varepsilon_0$ and (b) $\varepsilon_{\rm F}=0.5\varepsilon_0$. (c), (d) The PS allowed by energy conservation indicated by a yellow torus along with electron (red) and hole (blue) pockets in the low frequency domain for (c) $\varepsilon_{\rm F}=0.1\varepsilon_0$ and (d) $\varepsilon_{\rm F}=0.5\varepsilon_0$. 
(e), (f) Cross-sectional views of the electron-hole pockets and the PS allowed by energy conservation in the $k_x-k_z$ plane at frequencies where kinks appear in the optical conductivity.
}
\label{fig:optical_conductivity_Ef}
\end{figure}

{\em Optical conductivity with $\varepsilon_{\rm F}\neq0$.} ---
Next, we consider the case where $\varepsilon_{\rm F}\neq0$ in the presence of tilt. Figures \ref{fig:optical_conductivity_Ef}(a) and \ref{fig:optical_conductivity_Ef}(b) show calculated optical conductivities of NLSMs for $\varepsilon_{\rm F}=0.1\varepsilon_0$ and $\varepsilon_{\rm F}=0.5\varepsilon_0$, respectively, with $\Delta_{\rm t}=0.3 \varepsilon_0$. At low frequencies, the optical conductivity for $\varepsilon_{\rm F}=0.1\varepsilon_0$ increases linearly with increasing $\omega$, whereas that for $\varepsilon_{\rm F}=0.5\varepsilon_0$ exhibits an optical gap.

To address the difference in the low-frequency behaviors, in Figs. \ref{fig:optical_conductivity_Ef}(c) and \ref{fig:optical_conductivity_Ef}(d) we show the PS allowed by energy conservation corresponding to a low-frequency range along with electron-hole pockets. For $\varepsilon_{\rm F}=0.1\varepsilon_0$, there is an available PS for interband transitions consisting of two separate local domains located around the contact points between the electron and hole pockets. Similarly as we did for $\varepsilon_{\rm F}=0$, after replacing $\left|\Delta_{\rm t}\cos{\phi}\right|$ with $\left|\varepsilon_{\rm F}-\Delta_{\rm t}\cos{\phi}\right|$ in Eq.~(\ref{eq:NLSMs_optical_conductivity_notilt}),  we obtain the low-frequency optical conductivity as ${\sigma_{ii}\over \sigma_0} \approx C_i\frac{\hbar\omega}{\pi \Delta_{\rm t}}$ 
where $i=x,y,z$, $\phi_0=\cos^{-1}\left(\frac{\varepsilon_{\rm F}}{\Delta_{\rm t}}\right)$ is the location of graphene sheets sitting around the PS for interband transitions,
$C_x=\frac{\left(\frac{\varepsilon_{\rm F}}{\Delta_{\rm t}}\right)^2}{\sqrt{1-\left(\frac{\varepsilon_{\rm F}}{\Delta_{\rm t}}\right)^2}}$,
$C_y=\sqrt{1-\left(\frac{\varepsilon_{\rm F}}{\Delta_{\rm t}}\right)^2}$, and
$C_z=\frac{1}{\sqrt{1-\left(\frac{\varepsilon_{\rm F}}{\Delta_{\rm t}}\right)^2}}$.
Note that in the presence of finite $\varepsilon_{\rm F}$, the linear term dominates over the cubic one in the optical conductivity along the tilt direction, in contrast to the $\varepsilon_{\rm F}=0$ case [see Eq.~(\ref{eq:NLSMs_optical_conductivity_notilt})]. 
(For $\varepsilon_{\rm F}=\Delta_{\rm t}$ case, see Sec.~III in SM.)

For $\varepsilon_{\rm F}=0.5\varepsilon_0$, there is no available PS for interband transitions at low frequencies because the electron pocket becomes large enough to cover the entire PS allowed by energy conservation, leading to an optical gap due to Pauli blocking [see Fig.~\ref{fig:optical_conductivity_Ef}(d)]. The optical gap persists up to the frequency where the PS allowed by energy conservation touches the boundary of the electron pocket.
Similarly as in the $\varepsilon_{\rm F}=0$ case, we can obtain the size of the optical gap to be $\hbar\omega_{\rm gap}=2\frac{\varepsilon_{\rm F}-\Delta_{\rm t}}{\varepsilon_0+\Delta_{\rm t}} \varepsilon_0$, and thus the condition for the existence of an optical gap: $\varepsilon_{\rm F}>\Delta_{\rm t}$.

As $\varepsilon_{\rm F}$ increases, the electron (hole) pocket grows (shrinks), because the number of electrons in the system increases. The imbalanced sizes of the electron and hole pockets lead to two more kinks compared with the $\varepsilon_{\rm F}=0$ case, as shown in Figs.~\ref{fig:optical_conductivity_Ef}(e) and \ref{fig:optical_conductivity_Ef}(f). Similarly as in the $\varepsilon_{\rm F}=0$ case, we can obtain
$\hbar\omega_1=2\left|\frac{\varepsilon_{\rm F}-\Delta_{\rm t}}{\varepsilon_0+\Delta_{\rm t}}\right|\varepsilon_0$,
$\hbar\omega_2=2\left|\frac{\varepsilon_{\rm F}-\Delta_{\rm t}}{\varepsilon_0-\Delta_{\rm t}}\right|\varepsilon_0$,
$\hbar\omega_3=2\left|\frac{\varepsilon_{\rm F}+\Delta_{\rm t}}{\varepsilon_0+\Delta_{\rm t}}\right|\varepsilon_0$, and
$\hbar\omega_4=2\left|\frac{\varepsilon_{\rm F}+\Delta_{\rm t}}{\varepsilon_0-\Delta_{\rm t}}\right|\varepsilon_0$.
It follows that the condition for the existence of a flat region can be obtained from $\hbar\omega_4>\hbar\omega_5\equiv 2\varepsilon_0$, leading to  $\Delta_{\rm t}<\frac{1}{2}(\varepsilon_0-\varepsilon_{\rm F})$  \cite{supplemental}. Here $\omega_5$ is the frequency where the PS allowed by energy conservation changes its geometry from a torus to a spherelike manifold.

\begin{figure}[!htb]
\includegraphics[width=1\linewidth]{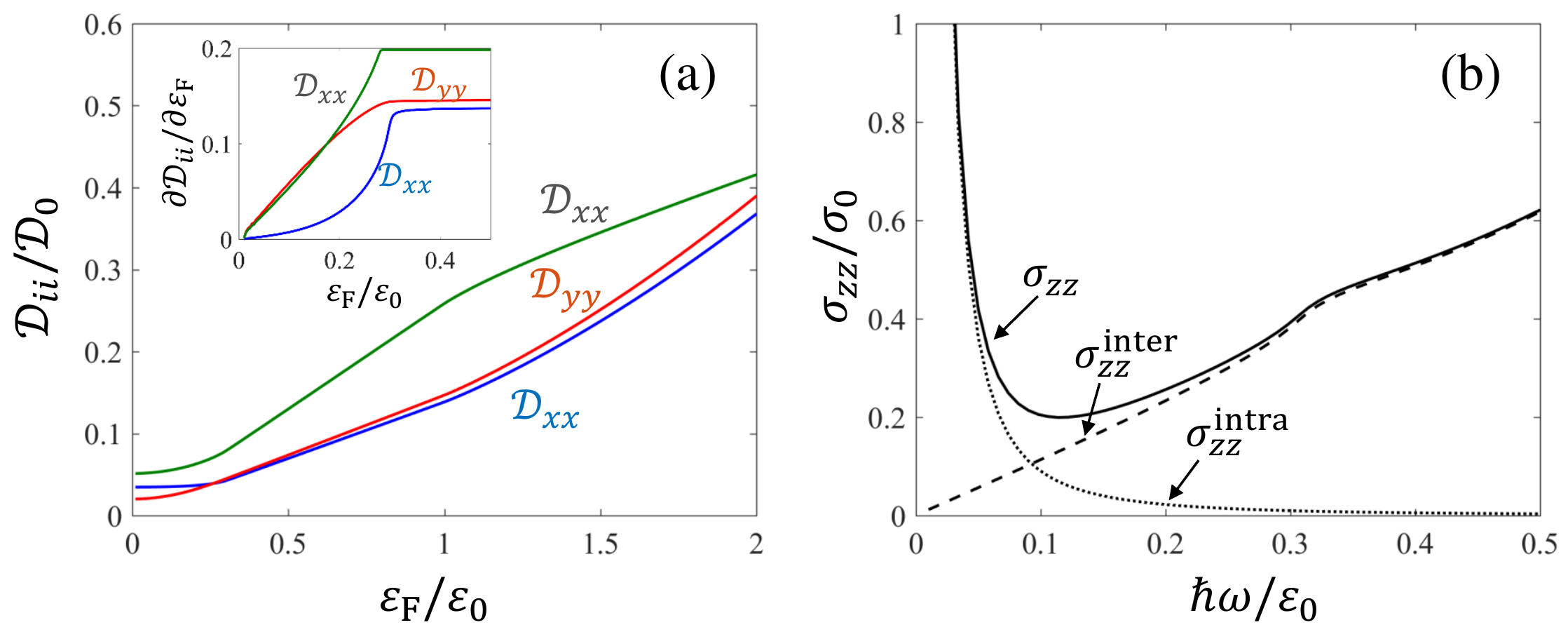}
\caption{
(a) The Drude weight of tilted NLSMs with $\Delta_{\rm t}=0.3\varepsilon_0$ and (b)
the intraband ($\sigma_{zz}^{\rm intra}$, dotted line) and interband ($\sigma_{zz}^{\rm inter}$, dashed line) contributions to the optical conductivity $\sigma_{zz}$ (solid line).
The inset to (a) shows the derivative of the Drude weight near $\varepsilon_{\rm F}=\Delta_{\rm t}$, where the Drude weight exhibits a nonanalytic kink behavior due to an abrupt change in the geometry of the FS. Here ${\cal D}_0=\frac{e^2}{\hbar}k_0\varepsilon_0$ and the Drude weight is defined to be $\sigma_{ii}^\mathrm{intra}={\cal D}_{ii}\delta(\hbar\omega)$. For (b), we use a finite broadening term $\eta=0.001\varepsilon_0$ replacing the $0^{+}$ term in Eq. (6) in SM.
}
\label{fig:Drude_weight}
\end{figure}

The intraband contribution to optical conductivity gives rise to a Drude peak at low frequencies whose weight also inherits a nonanalytic density dependence from the geometry of the FS. Interestingly, these are seen most clearly in the {\it derivatives} of the weight with respect to Fermi energy, as shown in Fig.~\ref{fig:Drude_weight}.

{\em Discussion.} ---
Introducing spin-orbit coupling (SOC) can gap out a nodal line and produce pairs of Weyl points \cite{Fang2015}. In such cases, at frequencies below the SOC scale, the PS for interband transitions are spheres enclosing the Weyl nodes, and the  conductivity will revert to linear frequency scaling known for Weyl semimetals (WSMs). Above the SOC energy, however, the separated PS recovers a toroidal shape, and this will have the characteristic dependence found in our work. Recent calculations for WSMs in the TaAs class indicate that these materials are weakly broken line node systems where the tilt scale dominates the SOC scale \cite{Bansil}. Thus, our analysis is applicable over a wide frequency range and can be used as a signature of these new states of matter in optical experiments.



\acknowledgments
We thank A. Bansil for communicating unpublished data on the band structure of WSMs in the TaAs family. This research was supported by the Basic Science Research Program through the National Research Foundation of Korea (NRF) funded by the Ministry of Education under Grant No. 2015R1D1A1A01058071. E.J.M.'s work on this project was supported by the U.S. Department of Energy, Office of Basic Energy Sciences under Award No. DE-FG02-ER45118. H.M. acknowledges travel support provided by the University Research Foundation at the University of Pennsylvania while this work was carried out.



\clearpage 
\widetext
\setcounter{section}{0}
\setcounter{equation}{0}
\setcounter{figure}{0} 
\setcounter{table}{0} 
\renewcommand\thefigure{S\arabic{figure}} 
\setcounter{page}{1}

\large
\begin{center}
{\bf Supplemental Material:\\
Electrodynamics on Fermi Cyclides in Nodal Line Semimetals}
\end{center}
\normalsize

\section{Toroidal coordinate transformation}
\label{sec:transformation}
In this section, we present a toroidal coordinate transformation for nodal line semimetals (NLSMs) and obtain the density of states (DOS) in the absence of tilt. Let us consider the following coordinate transformation
\begin{equation}\label{eq:transformation}
\begin{split}
k_x&\rightarrow (k_0+r\cos\theta) \cos\phi,\\
k_y&\rightarrow (k_0+r\cos\theta) \sin\phi,\\
k_z&\rightarrow r\sin\theta,
\end{split}
\end{equation}
which transforms Eq.~(1) in the main text to Eq.~(3) in the main text. Here, the Jacobian is given by $\mathcal{J}(r,\theta)=r(k_0+r\cos{\theta})$.

Using the toroidal coordinate transformation, consider the DOS in the absence of tilt ($\Delta_{t}=0$). Then the eigenenergies are given by $\varepsilon_{\pm}(r)=\pm \hbar v r$ and
the DOS can be easily obtained by
\begin{eqnarray}
    D(\varepsilon)
    &=&
    \frac{1}{(2\pi)^3}
    \int_0^{k_0}
    dr
    \int_0^{2\pi}
    d\theta
    \int_0^{2\pi}
    d\phi
    \mathcal{J}(r,\theta) \delta\left(\hbar v r - |\varepsilon|\right)\\
    &=&
    \frac{1}{(2\pi)^2}
    \int_0^{k_0}
    dr
    \int_0^{2\pi}
    d\theta \;
    r(k_0+r\cos{\theta})
    \frac{1}{\hbar v}\delta\left(r - \frac{|\varepsilon|}{\hbar v}\right)\\
    &=&
    \frac{k_0 |\varepsilon| }{2 \pi \hbar^2 v^2 },
\end{eqnarray}
which is valid only for $|\varepsilon|<\varepsilon_0\equiv\hbar v k_0$ where the toroidal structure is maintained.
Including the higher energy regime, the DOS is given by
\begin{equation}\label{eq:residual_conductivity}
{D(\varepsilon)\over D_0} =
\begin{cases}
{|\varepsilon|\over \varepsilon_0} & (|\varepsilon|\le \varepsilon_0), \\
{|\varepsilon|\over \varepsilon_0}\left[{1\over 2}+{1\over \pi}\sin^{-1}\left({\varepsilon_0\over |\varepsilon|}\right)
+{1\over \pi}\sqrt{\left({\varepsilon\over\varepsilon_0}\right)^2-1}
\right] & (|\varepsilon|> \varepsilon_0),
\end{cases}
\end{equation}
where $D_0={k_0\varepsilon_0 \over 2\pi\hbar^2 v^2}$.
Note that for $k_0\rightarrow 0$ or $\varepsilon\rightarrow \infty$, the DOS approaches that for Weyl semimetals given by $D(\varepsilon)={\varepsilon^2\over 2\pi^2\hbar^3 v^3}$.

\section{Kubo formula for optical conductivity}
\label{sec:Kubo_formula}

In this section, we present the Kubo formula for the optical conductivity and derive the low frequency asymptotic expressions directly from the Kubo formula. Here we show that to leading order in $\omega$, the optical conductivity obtained using the Kubo formula and that obtained by averaging optical conductivity of graphene sheets presented in the main text give the same result.

{\bf Optical conductivity.}
In the linear response and non-interacting limit, the optical conductivity can be calculated by using the Kubo Formula \cite{SM_Mahan}:
\begin{equation}
\label{eq:conductivity}
\sigma_{ij}(\omega)
=- \frac{ie^2}{\hbar} \sum_{s,s'} \int \frac{d^3 k}{(2\pi)^3} \frac{f_{s, \bm{k}}-f_{s',\bm{k}}}{\varepsilon_{s,\bm{k}}-\varepsilon_{s',\bm{k}}}
\frac{M^{ss'}_i(\bm k)M^{s's}_j(\bm k)}{\hbar\omega+\varepsilon_{s,\bm{k}}-\varepsilon_{s',\bm{k}}+i0^+},
\end{equation}
where $i,j=x,y,z$, $\varepsilon_{s,\bm{k}}$ and $f_{s,\bm{k}}=1/[1+e^{(\varepsilon_{s,\bm{k}}-\mu)/k_{\rm B}T}]$ are the eigenenergy and the Fermi distribution function for the band index $s=\pm$ and wave vector $\bm{k}$, respectively, $\mu$ is the chemical potential and $M^{ss'}_i(\bm k)=\langle{s,\bm{k}}|\hbar\hat{v}_i |{s',\bm{k}}\rangle$ with the velocity operator $\hat{v}_i=\frac{1}{\hbar}\frac{\partial \hat{H}}{\partial  k_i}$.
We can separate the real part of the longitudinal optical conductivity to the intraband and interband contributions:
\begin{equation}\label{eq:longitudinal_intra_before_transform}
\sigma^{\mathrm{intra}}_{ii}(\omega)=- \frac{\pi e^2}{\hbar} \sum_{s=\pm} \int \frac{d^3 k}{(2\pi)^3}\frac{\partial f_{s,\bm{k}}}{\partial \varepsilon_{s,\bm{k}}} |M^{ss}_i(\bm k)|^2\delta(\hbar\omega),
\end{equation}
\begin{equation}\label{eq:longitudinal_inter_before_transform}
\sigma^{\rm inter}_{ii}(\omega)
=-\frac{\pi e^2}{\hbar} \int \frac{d^3 k}{(2\pi)^3}
\frac{f_{-,\bm{k}}-f_{+,\bm{k}}}{\varepsilon_{-,\bm{k}}-\varepsilon_{+,\bm{k}}} |M^{-+}_i(\bm k)|^2
\delta(\hbar\omega+\varepsilon_{-,\bm{k}}-\varepsilon_{+,\bm{k}}).
\end{equation}

In the following we obtain the optical conductivity of NLSMs in the absence of tilt at zero Fermi energy. We show that for $\hbar\omega<2\varepsilon_0$ the results are in agreement with those obtained by averaging graphene optical conductivities presented in the main text, whereas for $\hbar\omega>2\varepsilon_0$ the optical conductivity of NLSMs shows linear frequency dependence as Weyl semimetal at high frequencies. In cylindrical coordinates ($k_x\rightarrow\rho \cos{\phi}$, $k_y\rightarrow\rho \sin{\phi}$), the eigenenergies of NLSMs are given by
$\varepsilon_\pm(\rho,\phi,k_z)=\pm \hbar v \sqrt{k_z^2 + (\rho-k_0)^2}$ and the interband matrix elements are calculated to be
\begin{eqnarray}
M_{x}^{-+}(\theta,\phi)&=&-i\hbar v\frac{k_z  \cos\phi}{\sqrt{k_z^2+(\rho-k_0)^2}},\\
M_{z}^{-+}(\theta,\phi)&=&-i\hbar v\frac{\rho-k_0 }{\sqrt{k_z^2+(\rho-k_0)^2}}.
\end{eqnarray}
By inserting these into Eq.~(\ref{eq:longitudinal_inter_before_transform}), we obtain
\begin{eqnarray}
    \sigma^\mathrm{inter}_{xx}(\omega)
    &=&
    \frac{e^2}{16\pi^2\hbar}
    \int_0^\infty d\rho \rho \int_0^{2\pi} d\phi \int_{-\infty}^\infty dk_z
    \frac{\hbar v k_z^2 \cos^2{\phi}}{[k_z^2+(\rho-k_0)^2]^{\frac{3}{2}}}
    \delta\left(\hbar\omega-2\hbar v \sqrt{k_z^2 + (\rho-k_0)^2}\right)\nonumber\\
    &=&
    \frac{1}{2}\sigma_0
    \left\{
        \Theta(2-\widetilde{\omega}) +
        \Theta(\widetilde{\omega}-2)
        \left[
            \frac{1}{2} + \frac{1}{3\pi}\sqrt{\widetilde{\omega}^2-4}
            \left(
                1+\frac{2}{\widetilde{\omega}^2}
            \right)
            +
            \frac{1}{\pi}
            \tan^{-1}\left(\frac{2}{\sqrt{\widetilde{\omega}^2-4}}\right)
        \right]
    \right\},
\end{eqnarray}
and
\begin{eqnarray}
    \sigma^\mathrm{inter}_{zz}(\omega)
    &=&
    \frac{e^2}{16\pi^2\hbar}
    \int_0^\infty d\rho \rho \int_0^{2\pi} d\phi \int_{-\infty}^\infty dk_z
    \frac{\hbar v (\rho-1)^2}{[k_z^2+(\rho-k_0)^2]^{\frac{3}{2}}}
    \delta\left(\hbar\omega-2\hbar v \sqrt{k_z^2 + (\rho-k_0)^2}\right)\nonumber\\
    &=&
    \sigma_0
    \left\{
        \Theta(2-\widetilde{\omega}) +
        \Theta(\widetilde{\omega}-2)
        \left[
            \frac{1}{2} + \frac{2}{3\pi}\sqrt{\widetilde{\omega}^2-4}
            \left(
                1-\frac{1}{\widetilde{\omega}^2}
            \right)
            +
            \frac{1}{\pi}
            \tan^{-1}\left(\frac{2}{\sqrt{\widetilde{\omega}^2-4}}\right)
        \right]
    \right\},
\end{eqnarray}
where $\sigma_0=\frac{e^2k_0}{16\hbar}$ and $\widetilde{\omega}=\hbar\omega/\varepsilon_0$.
Note that the optical conductivity is constant for $\hbar\omega<2\varepsilon_0$, and begins to increase at $\hbar\omega=2\varepsilon_0$. In the high frequency limit, $\sigma^\mathrm{inter}_{xx}(\omega)\approx\frac{e^2}{96\hbar v}\omega$ and $\sigma^\mathrm{inter}_{zz}(\omega)\approx\frac{e^2}{24\hbar v}\omega$, showing linear behaviors as Weyl semimetals.

\section{Low frequency asymptotic forms in the presence of tilt and Fermi energy}
\label{sec:asymptotic_forms}
Here we show derivations of the low frequency analytical expressions of the optical conductivity in the presence of tilt and Fermi energy. For the numerical calculation presented in the main text, we employed cylindrical coordinates that allow us to obtain valid expressions for the optical conductivity at all frequency range. For obtaining low frequency asymptotic expressions, however, it is not convenient to use cylindrical coordinates because the band structure appearing in the low-energy regime has a torus-like shape. Thus, we use the toroidal coordinates introduced in Sec.~\ref{sec:transformation} for the derivation of low frequency asymptotic forms.

The eigenenergies and eigenstates of the Hamiltonian of NLSMs in toroidal coordinates [Eq. (3) in the main text] are given by $\varepsilon_\pm(r,\theta,\phi)=\pm \varepsilon_0 \tilde{r} + \Delta_{t} (1 + \tilde{r} \cos{\theta}) \cos{\phi}$ where $\tilde{r}=r/k_0$ and $|\pm,\bm{k}\rangle=\frac{1}{\sqrt{2}}(\pm1,e^{i\phi})^\mathrm{T}$, respectively. The velocity matrices are expressed as $\hat{v}_x=\cos{\phi}\;\sigma_x$, $\hat{v}_y=\sin{\phi}\;\sigma_x$, and $\hat{v}_z=\sigma_y$.  It follows that the interband matrix elements used in the optical conductivity are
\begin{eqnarray}
M_{x}^{-+}(\theta,\phi)&=&-i\hbar v \sin{\theta} \cos{\phi},\\
M_{y}^{-+}(\theta,\phi)&=&-i\hbar v \sin {\theta} \sin {\phi},\\
M_{z}^{-+}(\theta,\phi)&=&i\hbar v \cos {\theta}.
\end{eqnarray}
By inserting these into Eq.~(\ref{eq:longitudinal_inter_before_transform}), we can obtain the interband part of the optical conductivity at zero temperature:
\begin{eqnarray}
\label{eq:longitudinal_inter_intermediate}
\sigma^{\rm inter}_{ii}(\omega)
&=&\frac{\pi e^2}{\hbar}
\int_0^{k_0} dr \int_0^{2\pi} d\theta \int_0^{2\pi} d\phi
\frac{\mathcal{J}(r,\theta)}{(2\pi)^3}
\frac{\mathcal{A}(\tilde{r},\theta,\phi)}
{2\varepsilon_0 \tilde{r}}
|M^{-+}_i(\theta,\phi)|^2
\delta(\hbar\omega-2\varepsilon_0 \tilde{r})\nonumber\\
&=&\frac{ e^2 k_0}{32\pi^2\hbar}
\int_0^{2\pi} d\theta \int_0^{2\pi} d\phi
\left(1+\frac{\hbar\omega}{2\varepsilon_0}\cos{\theta}\right)
\mathcal{A}\left(\frac{\hbar\omega}{2\varepsilon_0},\theta,\phi\right)
\left|\widetilde{M}_i^{-+}(\theta,\phi)\right|^2\nonumber\\
&=&\sigma_0 I_i(\omega),
\end{eqnarray}
where $\widetilde{M}_i^{-+}(\theta,\phi)=M_i^{-+}(\theta,\phi)/(\hbar v)$ and
$\mathcal{A}(\tilde{r},\theta,\phi)=\Theta \left[\varepsilon_{F}-\varepsilon_{-}(r,\theta,\phi)\right]-\Theta\left[\varepsilon_{F}-\varepsilon_{+}(r,\theta,\phi)\right]$
is the difference between the Fermi distribution functions (i.e., $f_{-,\bm{k}}-f_{+,\bm{k}}$) at zero temperature, which determines the integral area.
Here $\hbar\omega<2\varepsilon_0$ is assumed, and
\begin{eqnarray}
    I_i(\omega)
    &=&
{1\over 2\pi^2}
\int_0^{2\pi} d\theta \left(1+\frac{\hbar\omega}{2\varepsilon_0}\cos{\theta}\right)
\int_{0}^{2\pi} d\phi\,
\mathcal{A}\left(\frac{\hbar\omega}{2\varepsilon_0},\theta,\phi\right)
\left|\widetilde{M}_i^{-+}(\theta,\phi)\right|^2\nonumber\\
&=&
{1\over 2\pi^2}
\int_0^{2\pi} d\theta \left(1+\frac{\hbar\omega}{2\varepsilon_0}\cos{\theta}\right)
\left[\int_{\phi_{-}(\theta,\omega)}^{\phi_{+}(\theta,\omega)} d\phi
\left|\widetilde{M}_i^{-+}(\theta,\phi)\right|^2
+
\int_{-\phi_{+}(\theta,\omega)}^{-\phi_{-}(\theta,\omega)} d\phi
\left|\widetilde{M}_i^{-+}(\theta,\phi)\right|^2
\right]\nonumber\\
&=&
{1\over \pi^2}
\int_0^{2\pi} d\theta \left(1+\frac{\hbar\omega}{2\varepsilon_0}\cos{\theta}\right)
\int_{\phi_{-}(\theta,\omega)}^{\phi_{+}(\theta,\omega)} d\phi
\left|\widetilde{M}_i^{-+}(\theta,\phi)\right|^2,
\end{eqnarray}
where $\phi_\pm(\omega)=\cos^{-1}\left[
\frac{\varepsilon_{F}\mp\frac{\hbar\omega}{2}}
{\Delta_{t}\left(1+\frac{\hbar\omega}{2\varepsilon_0}\cos{\theta}\right)}\right]$. In the last line, we used $\left|\widetilde{M}_i^{-+}(\theta,\phi)\right|^2=\left|\widetilde{M}_i^{-+}(\theta,-\phi)\right|^2$.

Assuming $\varepsilon_{F}<\Delta_{t}$, we can expand $\phi_\pm(\theta,\omega)$ around $\omega=0$ as $\phi_\pm(\theta,\omega)\approx \phi_0+\Delta_{\pm}(\theta,\omega)$ where $\phi_0=\cos^{-1}\left(\frac{\varepsilon_{F}}{\Delta_{t}}\right)$ and
\begin{equation}\label{series_expansion}
\Delta_{\pm}(\theta,\omega)=\pm
    \frac{\varepsilon_0\pm \varepsilon_{F}\cos{\theta}}{2\sqrt{\Delta_{t}^2-\varepsilon_{F}^2}}
    \frac{\hbar\omega}{2\varepsilon_0}.
\end{equation}
Thus, at low frequencies $I_i(\omega)$ is calculated to be
\begin{eqnarray}
    I_i(\omega)
    &\approx&
{1\over \pi^2}\int_0^{2\pi} d\theta
\left(1+\frac{\hbar\omega}{2\varepsilon_0}\cos{\theta}\right)
\int_{\Delta_{-}(\theta,\omega)}^{\Delta_{+}(\theta,\omega)} d\phi
\left|\widetilde{M}_i^{-+}(\theta,\phi+\phi_0)\right|^2.
\end{eqnarray}
Note that the integral range of $\phi$ is $\Delta_{-}(\theta,\omega)<\phi<\Delta_{+}(\theta,\omega)$ and $\Delta_\pm(\theta,\omega)\propto\hbar\omega$ is small at low frequencies, thus the integrand can be expanded around $\phi=0$. For $I_x(\omega)$, $\left|\widetilde{M}_x^{-+}(\theta,\phi)\right|^2=\sin^2\theta \cos^2\phi$ thus we can obtain
\begin{eqnarray}
    I_x(\omega)
    &=&
{1\over \pi^2}\int_0^{2\pi} d\theta
\left(1+\frac{\hbar\omega}{2\varepsilon_0}\cos{\theta}\right)
\int_{\Delta_{-}(\theta,\omega)}^{\Delta_{+}(\theta,\omega)} d\phi\,
\sin^2{\theta}
\cos^2{(\phi+\phi_0)}\nonumber\\
&\approx&
{1\over \pi^2}\int_0^{2\pi} d\theta
\left(1+\frac{\hbar\omega}{2\varepsilon_0}\cos{\theta}\right)
\sin^2{\theta}
\int_{\Delta_{-}(\theta,\omega)}^{\Delta_{+}(\theta,\omega)} d\phi
\left[\cos^2{\phi_0}-\sin{(2\phi_0)}\phi-\cos{(2\phi_0)}\phi^2\right] \nonumber\\
&\approx&
\frac{ \left(\frac{\varepsilon_{F}}{\Delta_{t}}\right)^2}{\sqrt{1-\left(\frac{\varepsilon_{F}}{\Delta_{t}}\right)^2}}\frac{\hbar\omega}{\pi\Delta_{t}}.
\end{eqnarray}
Similarly, at low frequencies we can obtain the remaining components of the optical conductivity in units of $\sigma_0$ as
\begin{eqnarray}
I_y(\omega)
&\approx&
\sqrt{1-\left(\frac{\varepsilon_{F}}{\Delta_{t}}\right)^2}\frac{\hbar\omega}{\pi\Delta_{t}},
\\
I_z(\omega)
&\approx&
\frac{1 }
{\sqrt{1-\left(\frac{\varepsilon_{F}}{\Delta_{t}}\right)^2}}
\frac{\hbar\omega}{\pi\Delta_{t}}.
\end{eqnarray}
Thus the result obtained here to the leading order in $\omega$ is consistent with that in the main text obtained by averaging optical conductivity of graphene sheets.
Note that for $\varepsilon_{F}=0$, $\phi_0=\frac{\pi}{2}$ and thus $I_x(\omega)$ is written as
\begin{eqnarray}
I_x(\omega)
&\approx&
{1\over \pi^2}\int_0^{2\pi} d\theta
\left(1+\frac{\hbar\omega}{2\varepsilon_0}\cos{\theta}\right)
\sin^2{\theta}
\int_{\Delta_{-}(\theta,\omega)}^{\Delta_{+}(\theta,\omega)} d\phi \;
\phi^2 \nonumber\\
&\approx&
\frac{1}{12\pi^3}\left(\frac{\hbar\omega}{\Delta_{t}}\right)^3,
\end{eqnarray}
which is consistent with Eq.~(8) in the main text.

In the case of $\varepsilon_{F}=\Delta_{t}$, the series expansion described here fails for $I_{x}(\omega)$ and $I_{z}(\omega)$, thus the optical conductivity has a frequency dependence with a non-integer power. To obtain the power exponent, it is convenient to use Eq.~(6) in the main text, from which we can immediately obtain
\begin{eqnarray}
{\sigma_{ii} \over \sigma_0}
&\approx&
\int_{0}^{2\pi} {d\phi\over 2\pi}\,
\Theta({\hbar\omega-2\left|\varepsilon_{F}-\Delta_{t}\cos{\phi}\right|}) {\mathcal F}_{ii}(\phi)\nonumber\\
&=&
\int_{0}^{2\pi} {d\phi\over 2\pi}\,
\Theta({\hbar\omega-4\varepsilon_{F}\sin^2{\frac{\phi}{2}}}) {\mathcal F}_{ii}(\phi)\nonumber\\
&=&
2\int_{0}^{\phi_0} {d\phi\over 2\pi}\,
{\mathcal F}_{ii}(\phi),
\end{eqnarray}
where $\phi_0=2\sin^{-1}\sqrt{\frac{\hbar\omega}{4\varepsilon_{F}}}\approx\sqrt{\frac{\hbar\omega}{\varepsilon_{F}}}$. Thus, to the leading order in $\omega$, the optical conductivities for $\varepsilon_{F}=\Delta_{t}$ are calculated to be $\sigma_{xx}=\sigma_{zz}\approx\frac{\sigma_0}{\pi}\sqrt{\frac{\hbar\omega}{\varepsilon_{F}}}$ and $\sigma_{yy}\approx \frac{\sigma_0}{3\pi}\left(\frac{\hbar\omega}{\varepsilon_{F}}\right)^\frac{3}{2}$.

\section{Phase diagram for the existence of an optical gap and a flat region}
\label{sec:Phase_diagram}

\begin{figure}[!htb]
\includegraphics[width=0.8\linewidth]{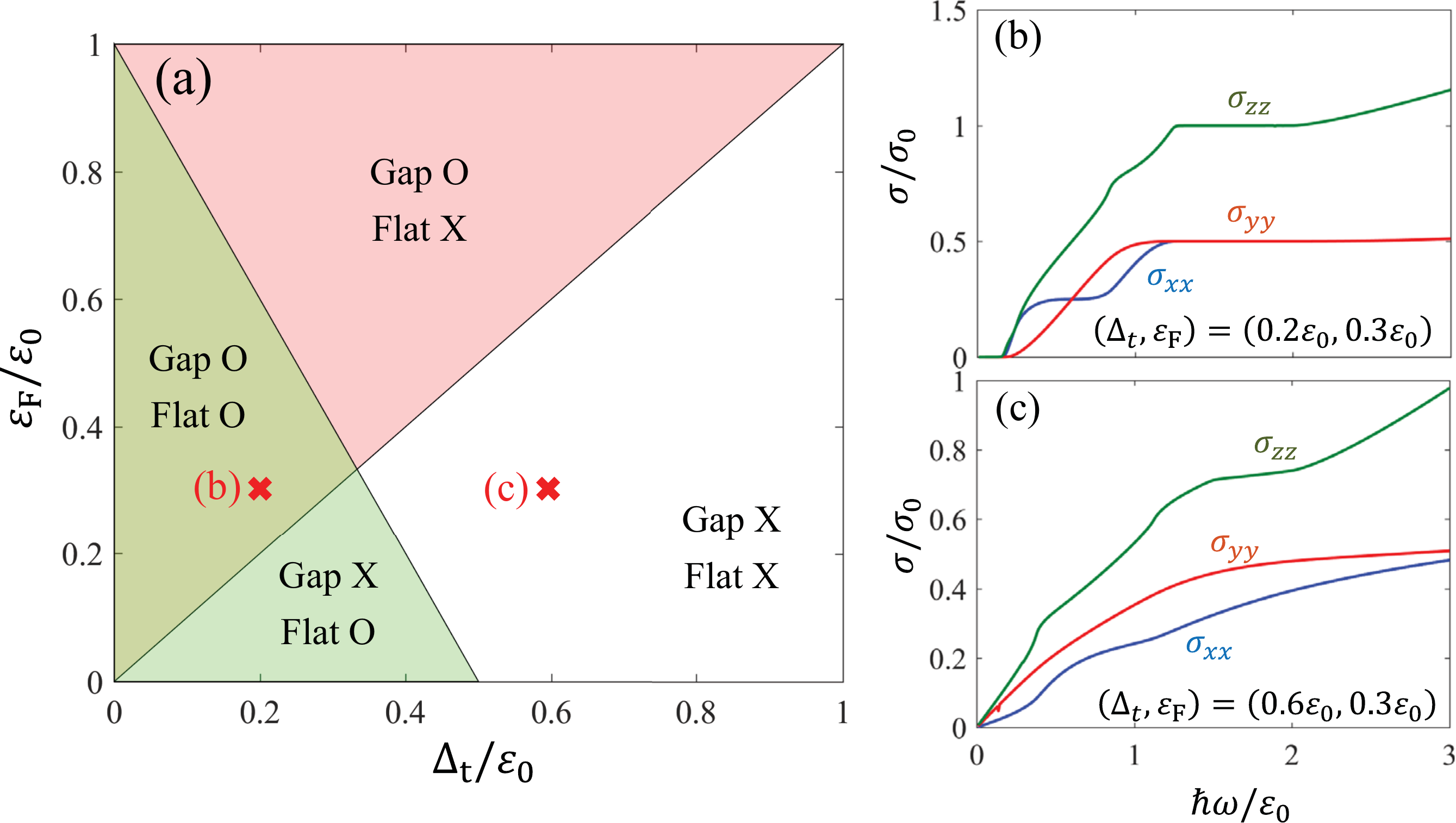}
\caption{
(a) Phase diagram for the existence of an optical gap in all components of optical conductivity and a flat region in $\sigma_{zz}$. (b), (c) Optical conductivities corresponding to the cross marks indicated in (a). Note that in (b), both the gap and flat region are present, whereas in (c), both are absent.
}
\label{fig:phase_diagram}
\end{figure}

The condition for the appearance of an optical gap in all components of optical conductivity ($\varepsilon_{F}>\Delta_{t}$) and that of a flat region in $\sigma_{zz}$ at intermediate frequencies [$\Delta_{t}<{1\over 2}(\varepsilon_0-\varepsilon_{F})$] are independent of each other, creating four possible scenarios where an optical gap is present/absent at low frequencies with/without a flat region. Figure \ref{fig:phase_diagram} (a) shows the corresponding phase diagram on the $\Delta_{t}/\varepsilon_0$  and $\varepsilon_{F}/\varepsilon_0$ plane. Figures \ref{fig:phase_diagram} (b) and (c) show the remaining cases not treated in the main text, where both the gap and flat region are present or absent.



\section{Comparison with other models}
\label{sec:model_comparison}
\subsection{4-band continuum model}

\begin{figure}[!htb]
\includegraphics[width=0.8\linewidth]{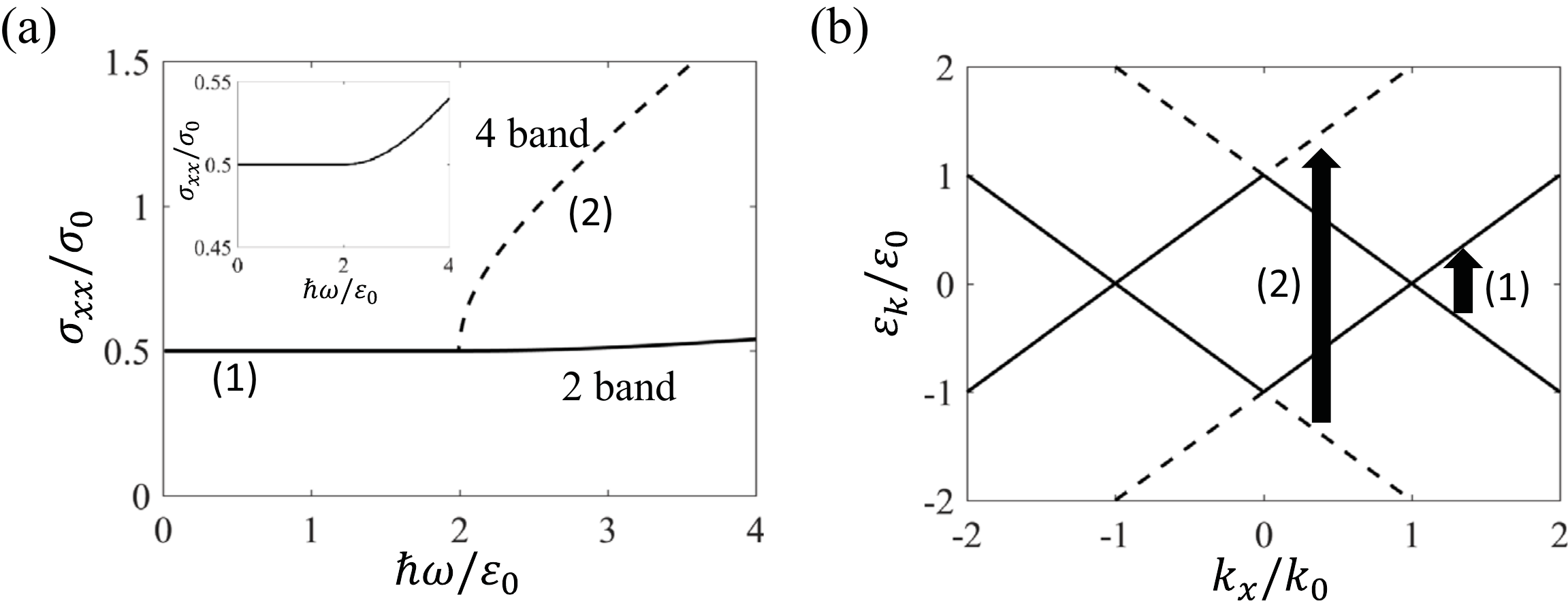}
\caption{
(a) Optical conductivities for the 2-band (solid)  and 4-band (dashed) models in the absence of tilt with zero Fermi energy. (b) Energy dispersions along the $k_x$ axis for the 2-band (solid) and the 4-band (dashed) models. The thick arrows labeled by (1) and (2) represent interband transitions at low frequencies and high frequencies, respectively. The inset to (a) shows an enlarged view in $\sigma_{xx}$ near $\hbar\omega=2\varepsilon_0$ for the 2-band model.
}
\label{fig:model_comparison_4band}
\end{figure}

In this section, we compare the optical conductivities obtained from the 2-band and 4-band continuum models, and demonstrate that in the frequency range where interband transitions involving higher energy bands are negligible, they give the same results.

Consider the following 4-band continuum model \cite{SM_Burkov2011,SM_Koshino2016}:
\begin{equation}\label{eq:hamiltonian_4band}
H=\hbar v (k_x\sigma_x+k_y\sigma_y) + \hbar v k_z \tau_z \sigma_z+\varepsilon_0 \tau_x +\hbar \bm{v}_{t}\cdot\bm{k}\sigma_0.
\end{equation}
The eigenenergies are given by
\begin{equation}
\varepsilon_{\pm}^2=\hbar^2v^2\left[(k_{\rho}\pm k_0)^2 + k_z^2\right] +\hbar \bm{v}_{t}\cdot\bm{k},
\end{equation}
where $\varepsilon_0=\hbar v k_0$. Thus, $\varepsilon_{-}$ has zero energy solutions 
forming a nodal line with the radius of $k_0$. Note that the energy dispersion for the low-energy bands exactly match with that of the 2-band model
[see Fig.~\ref{fig:model_comparison_4band} (b)].

Next, consider optical conductivities for the 2-band and 4-band models in the absence of tilt with zero Fermi energy, for simplicity [see Fig.~\ref{fig:model_comparison_4band} (a)].
At frequencies $\hbar\omega<2\varepsilon_0$, the results for both models are in good agreement because the two models have the same low-energy electronic structure. This is true when NLSMs are tilted; a tilt term changes the shape of the band dispersion but not the wave functions, and thus there is no mixing between low and high energy states.
At $\hbar\omega>2\varepsilon_0$, however, the 4-band result strongly deviates from the 2-band result, showing a sharp increase at $\hbar\omega=2\varepsilon_0$ due to interband transitions between high energy bands that are not captured by the 2-band model \cite{SM_Carbotte2017}.
Thus in the frequency range where interband transitions involving high energy bands are negligible, the optical conductivity of NLSMs shows the exactly same qualitative features as described in the main text, despite the existence of the high energy bands that are not captured by the 2-band model.

\subsection{2-band lattice model}

In this section, we present the optical conductivities obtained from a 2-band lattice model, and demonstrate that at low frequencies, the result is consistent with the corresponding 2-band continuum model.

\begin{figure}[!htb]
\includegraphics[width=0.8\linewidth]{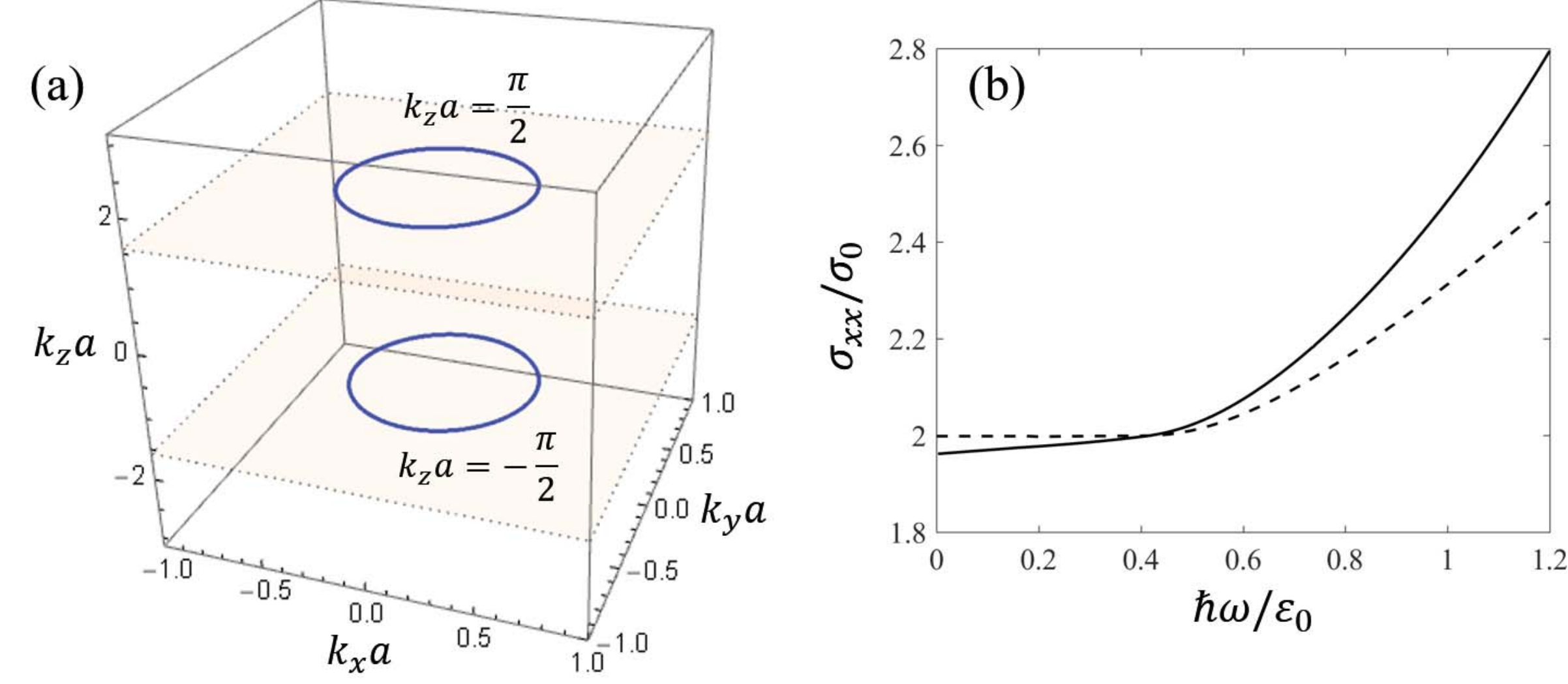}
\caption{
(a) Nodal lines in the first Brillouin zone for $m_0=3.8$. (b) Optical conductivities $\sigma_{xx}$ for the 2-band lattice (solid) and continuum (dashed) models in the absence of tilt with zero Fermi energy and $m_0=3.8$.  Here $\sigma_0=\frac{e^2 k_0}{16\hbar}\frac{v}{v_z}$ and $t=t_z=\hbar v/a$ was used for the calculation. Note that two nodal lines contribute to the optical conductivity and thus $\sigma_{xx}$ approaches $2\sigma_0$ as $\omega\rightarrow 0$.
}
\label{fig:model_comparison_lattice}
\end{figure}

Consider the following 2-band lattice model:
\begin{equation}
\label{eq:lattice}
H=t\{m_0-2[\cos(k_x a)+\cos(k_y a)]\}\sigma_x + t_z \cos(k_z a) \sigma_y,
\end{equation}
where $a$ is the lattice spacing, and $t$, $t_z$ and $m_0$ are material-dependent parameters. For $0<m_0<4$, the Hamiltonian in Eq.~(\ref{eq:lattice}) shows two nodal lines on $k_z=\pm {\pi\over 2a}$ planes centered at $k_x=k_y=0$ in the first Brillouin zone [see Fig.~\ref{fig:model_comparison_lattice}(a)]. The nodal lines touches the $k_x$ and $k_y$ axes at $k_x=\pm k_0$ and $k_y=\pm k_0$, respectively, 
where $m_0=2[\cos(k_0 a)+1]$. For simplicity, assume that $k_0 a\ll 1$. Then $m_0\approx 4-(k_0 a)^2$ and the low-energy continuum Hamiltonian for the nodal lines near $\bm{k}=(0,0,\pm {\pi\over 2a})$ has the form
\begin{equation}
\label{eq:2band_quadratic}
H=\frac{\hbar v}{k_0}\left(k_\rho^2-k_0^2\right)\sigma_x \mp \hbar v_zk_z\sigma_y,
\end{equation}
where ${\hbar v\over a}=t k_0 a$, ${\hbar v_z\over a}=t_z$ and $k_\rho=\sqrt{k_x^2+k_y^2}$.
Note that the band dispersion along the in-plane direction is quadratic, unlike the continuum model introduced in the main text [see Eq.~(1)] whose dispersion is linear along the in-plane direction. Similarly as in Sec.~\ref{sec:asymptotic_forms}, we can obtain the low-frequency optical conductivities as $\sigma_{xx}=\frac{e^2 k_0}{16\hbar}\frac{v}{v_z}$ and $\sigma_{zz}=\frac{e^2 k_0}{32 \hbar}\frac{v_z}{v}$ for $\omega<2v k_0$. 
This means that the Hamiltonian in Eq.~(28) has the same qualitative features with constant optical conductivities at low frequencies as that in Eq.~(1) in the main text.

Figure \ref{fig:model_comparison_lattice}(b) shows the calculated optical conductivities $\sigma_{xx}$ for the 2-band lattice and continuum models. At low frequencies, 
the two results are in good agreement. As the frequency increases, however, the optical conductivity of the lattice model begins to deviate from that of the continuum model because of contributions from the high-energy band structure that cannot be captured by the low-energy continuum model.

\section{Optical conductivity with different forms of tilt}
\subsection{Out-of-plane tilt}
\label{sec:kz_tilt}
\begin{figure}[!htb]
\includegraphics[width=0.8\linewidth]{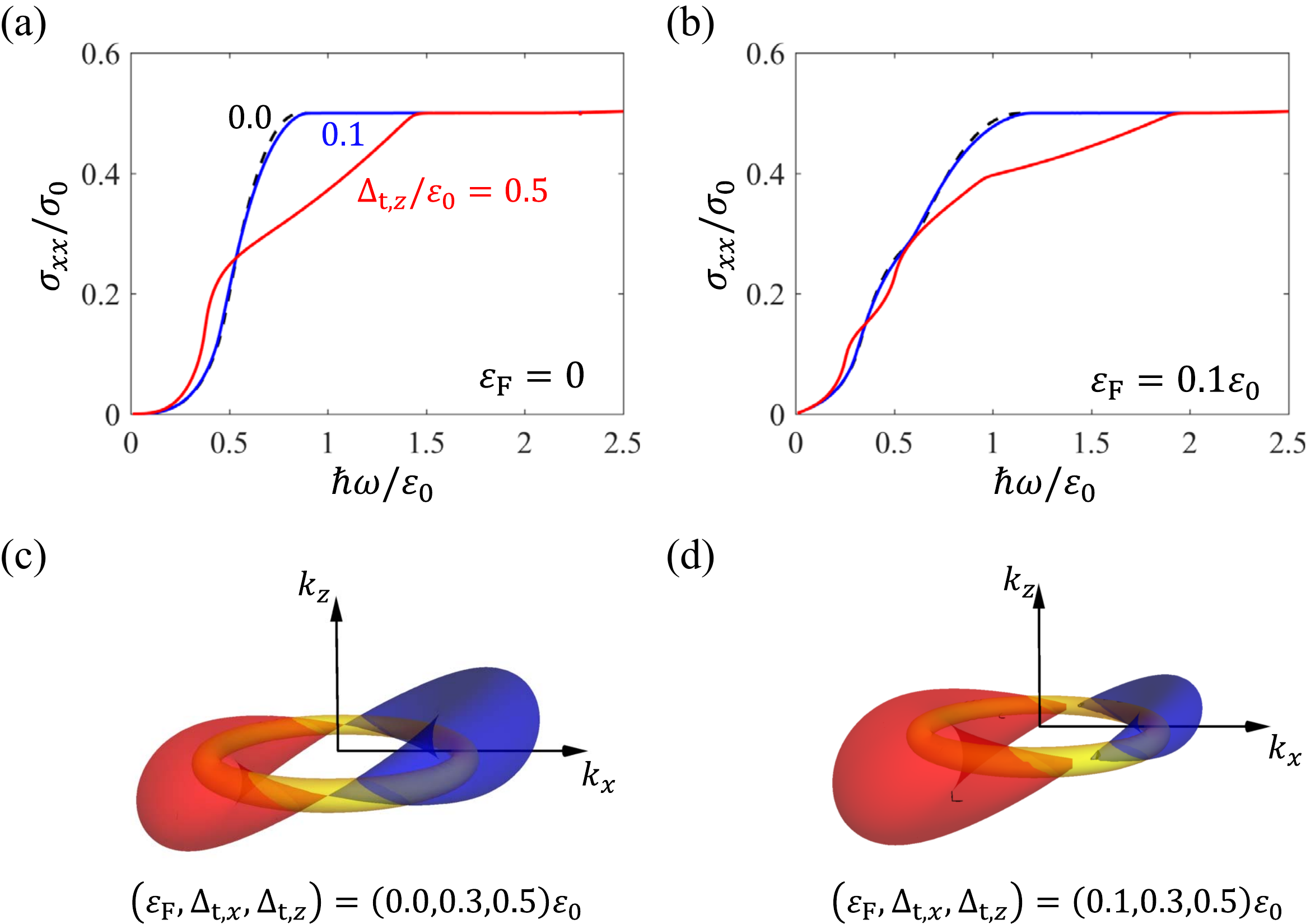}
\caption{
(a), (b) Optical conductivities with (a) $\varepsilon_{F}=0$ and (b) $\varepsilon_{F}=0.1 \varepsilon_0$ in the presence of both the in-plane ($\Delta_{{t},x}$) and out-of-plane ($\Delta_{{t},z}$) tilts. (c), (d) Electron and hole pockets exhibiting distorted Dupin cyclide due to the out-of-plane tilt. Note that the out-of-plane tilt just makes the FS asymmetric along the $k_z$-axis, without changing any essential features such as the point contact between the electron and hole pockets, thus leading to the same qualitative low and high frequency features in the optical conductivity compared with that obtained without the out-of-plane tilt.
}
\label{fig:kz_tilt}
\end{figure}
In this section, we consider the effect of the out-of-plane tilt, which is neglected for simplicity in the main text. Here we set the tilt velocity to be $\bm{v}_t=v_{t,x}\hat{x}+v_{t,z}\hat{z}$. The corresponding Hamiltonian describing the tilt can be written as $H_t=(\Delta_{t,x}\tilde{k}_x+\Delta_{t,z}\tilde{k}_z)\sigma_0$, where $\Delta_{t,i}=\hbar v_{t,i} k_0$ and $\tilde{k}_i=k_i/k_0$. Figures \ref{fig:kz_tilt} (a) and (b) show the optical conductivity $\sigma_{xx}$ with a fixed in-plane tilt energy $\Delta_{t,x}=0.3\varepsilon_0$ and various out-of-plane tilt energies $\Delta_{t,z}/\varepsilon_0=0.0,0.1,0.5$.
It is important to notice that at low frequencies the characteristic frequency dependence is robust against the change of $\Delta_{t,z}$; $\sigma_{xx}$ for $\varepsilon_{F}=0$ exhibits cubic frequency dependence while for $\varepsilon_{F}=0.1 \varepsilon_0$, $\sigma_{xx}$ shows linear behavior, which is consistent with our results in the main text where we assume $\Delta_{t,z}=0$.
As we increase the frequency, the optical conductivity with $\Delta_{t,z}\neq0$ begins to deviate from that with $\Delta_{t,z}=0$ because the distortion of the FS by the out-of-plane tilt leads to a different geometry of the PS allowed for interband transitions.
At high frequencies, however, the PS allowed for interband transitions fully covers the FS, leading to the same optical conductivity behavior regardless of the existence of tilt. Note that compared with the FS in the absence of the out-of-plane tilt (see Figs. 3 and 4 in the main text), the out-of-plane tilt simply distorts the electron and hole pockets making the FS asymmetric along the $k_z$-axis, but without changing any qualitative features such as the point contact between the electron and hole pockets, as shown in Figs.~\ref{fig:kz_tilt}(c) and (d).

\subsection{Fluctuating in-plane tilt}
\label{sec:arbitrary_tilt}
\begin{figure}[!htb]
\includegraphics[width=0.8\linewidth]{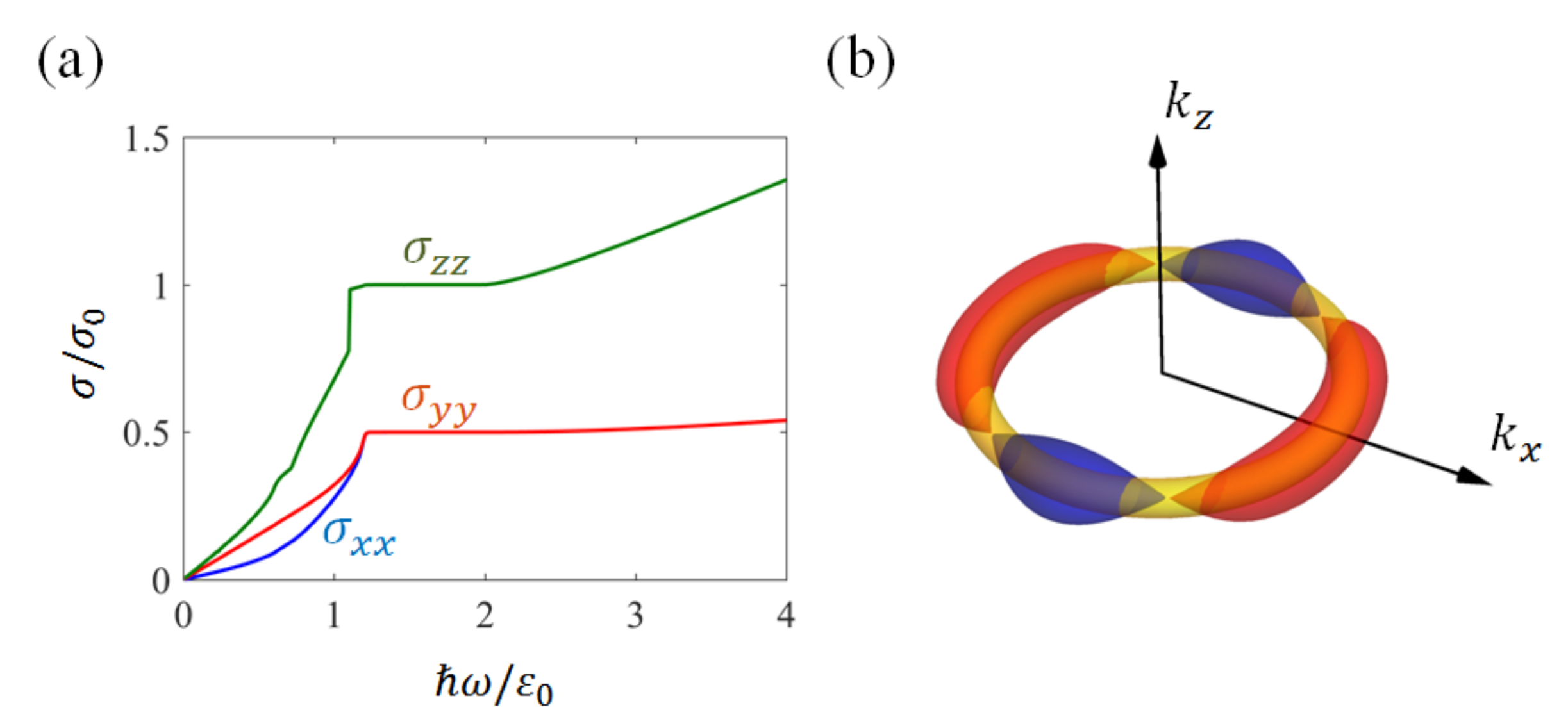}
\caption{
(a) The optical conductivity in the presence of a fluctuating tilt and (b) the corresponding electron and hole pockets which meet at multiple points. Here we used a tilt term $H_{t}=\Delta_{t} \cos{\left(\frac{3k_x}{k_0}\right)}\sigma_0$ and $\Delta_{t}=0.6\varepsilon_0$.
}
\label{fig:arbitrary_tilt}
\end{figure}

Here we consider the case where the tilt term fluctuates along the nodal line so that the electron and hole pockets meet at multiple points, as shown in Fig.~\ref{fig:arbitrary_tilt} (b). In such cases, the FS does not exhibit the Dupin cyclide geometry described in the main text, yet the underlying physics is essentially the same as the case where only a linear tilt exists. Note that at low frequencies, the PS allowed for interband transitions have a similar geometry as the case where only a linear tilt exists, except for the number of the contact points [see Fig.~\ref{fig:arbitrary_tilt} (b)]. In the presence of multiple contact points, the total conductivity at low frequencies is obtained by summing up all the contributions from each contact point. Note that each contact point can show either linear or cubic scaling behavior at low frequencies depending on the location and the conductivity direction with respect to the tilt, as shown in the main text. If linear and cubic scaling behaviors coexist, the linear behavior becomes dominant over the cubic behavior in the zero frequency limit. In Fig.~\ref{fig:arbitrary_tilt} (b), the contact points are located away from the $k_x=0$ plane, and thus optical conductivities contributed from each contact point show linear behavior at low frequencies for $\sigma_{xx}$. This leads to a linear behavior of $\sigma_{xx}$ at low frequencies, as shown in Fig.~\ref{fig:arbitrary_tilt} (a).



\end{document}